\documentclass[11pt,a4paper]{article}
\usepackage{amsmath}
\usepackage[mathcal]{eucal}
\usepackage{latexsym}
\usepackage{amssymb}
\begin{titlepage}
\title{Instanton representation of Plebanski gravity. Application to gravitons about DeSitter spacetime}
\author{Eyo Eyo Ita III}
\medskip
\input amssym.def
\input amssym.tex

\def \in{\indent}

\begin{document}
\maketitle
\bigskip
\centerline{Physics Department} 
\smallskip
\centerline{United States Naval Academy}
\smallskip
\centerline{572c Holloway Road. Annapolis, Maryland 21402}
\smallskip
\centerline{ita@usna.edu} 
  
\bigskip   
                   
\begin{abstract}
Using the instanton representation method, we re-construct graviton solutions about DeSitter spacetime.  We have used this example as a testing arena to expose the internal structure of the method and to establish that it works for known solutions.  This paper is a precursor for its application to the construction of new General Relativity solutions in future work.
\end{abstract}
\end{titlepage}

\section{Introduction and background}

In \cite{EYOITA} a new formulation of gravity has been presented referred to as the instanton representation of Plebanski gravity.  It was shown that the instanton representation is related to the Ashtekar formalism \cite{ASH2} by the exchange of a certain variable.  The Instanton representation and the Ashtekar formulations of gravity are both complementary, in that they can be seen as two daughter theories arising from the same mother theory, namely the Plebanski theory of gravity (See e.g. \cite{EYOITA}).  The associated action for the instanton representation $I_{Inst}$ can be written in 3+1 form as
\begin{eqnarray}
\label{THEACTION}
I_{Inst}=\int{dt}\int_{\Sigma}d^3x\biggl[\Psi_{ae}B^i_a\dot{A}^a_i+A^a_0B^i_eD_i\Psi_{ae}\nonumber\\-\epsilon_{ijk}N^iB^j_aB^k_e
-{i \over 2}N\sqrt{\hbox{det}B}\sqrt{\hbox{det}\Psi}\bigl(\Lambda+\hbox{tr}\Psi^{-1}\bigr)\biggr],
\end{eqnarray}
where $\Sigma$ is a 3-dimensional spatial hypersurface embedded in a 4-dimensional spacetime of topology $M=\Sigma\times{R}$.  The basic variables are a gauge connection $A^a_{\mu}=(A^a_0,A^a_i)$ and a 3 by 3 matrix $\Psi_{ae}$, which take their values in the special orthogonal group in three dimensions $SO(3,C)$,\footnote{For index conventions we use lower case symbols from the beginning of the Latin alphabet $a,b,c,\dots$ to denote internal $SO(3,C)$ indices, and from the middle $i,j,k,\dots$ for spatial indices.  Spacetime indices are denoted by Greek symbols $\mu,\nu,\dots$.} the set of complex 3-by-3 matrices $O$ such that $O^T=O^{-1}$ and $det{O}=1$.  Whereas in the Plebanski formulation $\Psi_{ae}$ is regarded as an auxiliary field, in the instanton representation $\Psi_{ae}$ is a full dynamical variable on equal footing with $A^a_i$.\par 
\indent  
The $SO(3,C)$ field strength of $A^a_{\mu}$ is given by 
\begin{eqnarray}
\label{FIRST1}
F^a_{\mu\nu}=\partial_{\mu}A^a_{\nu}-\partial_{\nu}A^a_{\mu}+f^{abc}A^b_{\mu}A^c_{\nu},
\end{eqnarray}
\noindent
of which $F^a_{0i}$ and $B^i_e={1 \over 2}\epsilon^{ijk}F^e_{jk}$ are respectively its temporal component and magnetic field.  The auxiliary fields $(A^a_0,N^i,N)$ are the temporal connection component, shift vector and lapse function whose variations yield the Gauss' law, vector and Hamiltonian constraints respectively\footnote{The constraints have also appeared in this form in \cite{CAP} within the context of the initial value problem of General Relativity in the CDJ formalism.}    
\begin{eqnarray}
\label{VALUE}
B^i_eD_i\{\Psi_{ae}\}=0;~~\epsilon_{dae}\Psi_{ae}=0;~~\Lambda+\hbox{tr}\Psi^{-1}=0,
\end{eqnarray}
\noindent
and $\Lambda$ is the cosmological constant.  Note that we must have $(\hbox{det}\Psi)\neq{0}$ and $(\hbox{det}B)\neq{0}$, a set of nondegeneracy conditions which limits the regime of equivalence between the instanton representation of Plebanski gravity and General Relativity to spacetimes of Petrov Types I, D, and O.  Given that the basic variables in general are complex for Lorentzian signature spacetimes, the action (\ref{THEACTION}) must additionally be supplemented by reality conditions which guarantee that the spacetime metric is real.  For the Euclidean signature case the reality conditions are automatically implemented by performing a Wick rotation $N\rightarrow{i}N$, and requiring all variables $A^a_i,\Psi_{ae}$ to be real.\par 
\indent  
The main result of this paper will be the construction of gravitons propagating on a Euclidean DeSitter background, using the instanton representation method based on (\ref{THEACTION}).  This solution is already known in the literature via conventional methods, and so the purpose of this paper will be to reproduce it using the instanton representation method in order to provide further evidence that the method works for known solutions.  As with any new method, it is important to establish first that it is capable of producing the standard solutions as a kind of test arena, which also lends some intuition and insight into the structure of the method.  Then with this insight, one can apply the method within a new context in the construction of solutions which may be new, an area of future research.  The instanton representation method has so far been tested in the time-independent case in \cite{EYOITA1}, leading to Schwarzschild-like solutions and a corroboration of Birkhoff's theorem.\par  
\indent 
Thus far we have provided some preliminaries on the group structure and dynamical variables.  In this section we will continue with the mathematical structure of the method.  This will provide the setting for the construction of solutions to the Einstein field equations.  A main problem in dealing with the Einstein equations in standard metric General Relativity for general spacetimes where no symmetry reductions are present, is the separation of physical from gauge effects (due to the coordinate system).  We will show that the instanton representation method enables one to have a clean separation of these degrees of freedom, and provides an interpretation of coordinate-induced effects of gravity within a Yang--Mills setting.

\subsection{Instanton representation equations of motion}

In addition to the intial value constraints, the instanton representation provides two equations of motion for the dynamical variables.  The initial value constraints (\ref{VALUE}) can always be imposed at the level after the dynamical equations have been written down, but not at the level of the action (\ref{THEACTION}).  The Lagrange's equation for $\Psi_{ae}$ is given by \cite{EYOITA}
\begin{eqnarray}
\label{FIRST}
B^i_eF^a_{0i}+(\hbox{det}B)N^i(B^{-1})^d_i\epsilon_{dae}+iN(\hbox{det}B)^{1/2}\sqrt{\hbox{det}\Psi}(\Psi^{-1}\Psi^{-1})^{ea}=0.
\end{eqnarray}
\noindent
It can be shown that the Lagrange's equation for $A^a_{\mu}$, once the vector and Hamiltonian constraints have been implemented, simplifies to \cite{EYOITA}
\begin{eqnarray}
\label{RECALL1}
\epsilon^{\mu\nu\rho\sigma}D_{\nu}(\Psi_{ae}F^e_{\rho\sigma})=0.
\end{eqnarray}
\noindent
In (\ref{RECALL1}), $D_{\nu}$ is the $SO(3,C)$ gauge covariant derivative with $SO(3)$ structure 
constants $f_{abc}=\epsilon_{abc}$ acting on $\Psi_{ae}$, seen as a second rank $SO(3,C)$ tensor  
\begin{eqnarray}
\label{FIRST2}
D_{\nu}\Psi_{ae}=\partial_{\nu}\Psi_{ae}+A^b_{\nu}\bigl(f_{abc}\Psi_{ce}+f_{ebc}\Psi_{ac}\bigr).
\end{eqnarray}
\noindent
Applying the Leibniz rule and the Bianchi identity to (\ref{RECALL1}), we have
\begin{eqnarray}
\label{RECALL2}
\epsilon^{\mu\nu\rho\sigma}F^e_{\rho\sigma}D_{\nu}\Psi_{ae}+\Psi_{ae}\epsilon^{\mu\nu\rho\sigma}D_{\nu}F^e_{\rho\sigma}=
\epsilon^{\mu\nu\rho\sigma}F^e_{\rho\sigma}D_{\nu}\Psi_{ae}=0.
\end{eqnarray}
\noindent
The $\mu=0$ component of (\ref{RECALL2}) yields $B^i_eD_i\Psi=0$ which is the Gauss' law constraint, the first equation 
of (\ref{VALUE}).\par  
\indent
Since $(\hbox{det}B)\neq{0}$, then we can multiply (\ref{FIRST}) by $(B^{-1})^e_i$ to get 
\begin{eqnarray}
\label{RECALLL}
F^a_{0i}+\epsilon_{ijk}B^j_aN^k+iN\sqrt{\hbox{det}B}\sqrt{\hbox{det}\Psi}(B^{-1})^d_i(\Psi^{-1}\Psi^{-1})^{ad}.
\end{eqnarray}
\noindent
Equation (\ref{RECALLL}) states that the gauge curvature of $A^a_{\mu}$ is Hodge self-dual with respect to a certain spacetime metric $g_{\mu\nu}$, and (\ref{RECALL1}) implies that this $g_{\mu\nu}$ solves the Einstein equations when the initial value constraints (\ref{VALUE}) hold \cite{EYOITA}.  To construct $g_{\mu\nu}$ one defines a quantity $c_{ij}$, given by
\begin{eqnarray}
\label{CONSTRU}
c_{ij}=F^a_{0i}(B^{-1})^a_j;~~c\equiv\hbox{det}(c_{(ij)}).
\end{eqnarray}
\noindent
Then spliting (\ref{CONSTRU}) into its symmetric and antisymmetric parts defines a spatial 3-metric $(h_{ij})_{Hodge}$ and a shift vector $N^i$ given by
\begin{eqnarray}
\label{CONSTRU1}
(h_{ij})_{Hodge}=-{{N^2} \over c}c_{(ij)};~~N^i=-{1 \over 2}\epsilon^{ijk}c_{jk},
\end{eqnarray}
\noindent
with the lapse function $N$ freely specifiable.  A spatial 3-metric can also be constructed directly from solutions to the initial value constraints (\ref{VALUE}), according to the formula
\begin{eqnarray}
\label{MEETRIC}
(h_{ij})_{Constraints}=(\hbox{det}\Psi)(\Psi^{-1}\Psi^{-1})^{ae}(B^{-1})^a_i(B^{-1})^e_j(\hbox{det}B).
\end{eqnarray}
\noindent
Equation (\ref{MEETRIC}) uses only the spatial connection $A^a_i$ and contains no reference to a shift vector $N^i$ or to time evolution, whereas (\ref{CONSTRU1}) involves velocities $\dot{A}^a_i$ through the $F^a_{0i}$ terms of (\ref{FIRST1}).  Equations (\ref{CONSTRU1}) and (\ref{MEETRIC}) feature spatial metrics $h_{ij}$ constructed according to two separate criteria, and as a consistency condition we demand that they be equal to each other
\begin{eqnarray}
\label{FIRST3}
(h_{ij})_{Constraints}=(h_{ij})_{Hodge}.
\end{eqnarray}
\noindent
Equation (\ref{FIRST3}) is the cornerstone of what we will refer to as the instanton representation method for constructing solutions to the Einstein equations.\par  
\indent
The constraint solutions can be classified according to the Petrov type of spacetime, which depends on the multiplicity of eigenvalues of $\Psi_{ae}$ (See e.g. \cite{MACCALLUM} and \cite{PENROSERIND}).  The condition (\ref{FIRST3}) places stringent constraints on the form of the metric, which appears to lead almost uniquely to the desired solution for the corresponding Petrov Type.   

\subsection{Organization of this paper}

The organization of this paper is as follows.  The background solution will be denoted by $(\Psi_{ae})_o$ and $(A^a_i)_o$, and their respective perturbations by $\epsilon_{ae}$ and $a^a_i$.  In section 2 we write the initial value constraints for a Petrov Type O background and in section 3 we linearize the contraints about this background using a homogeneous and isotropic connection.  We solve the constraints, obtaining the spin 2 configurations for the $\epsilon_{ae}$, and impose by hand an associated gauge-fixing condition on $a^a_i$.  This provides the ingredients for $(h_{ij})_{Constraints}$, which we explicitly construct.  In section 4 we construct $(h_{ij})_{Hodge}$ and impose the consistency condition (\ref{FIRST3}).  This fixes the background solution as DeSitter space, and provides an evolution equation for $a^a_i$ in terms of $\epsilon_{ae}$.  In section 5 we find the explicit time dependence of $\epsilon_{ae}$ using (\ref{RECALL1}), which in turn fixes $a^a_i$ and the 
3-metric $h_{ij}$ in section 6.  In section 7 we provide a summary of our results and a conclusion.\par
\indent 
There is a final note regarding indices in this paper.  We will often not distinguish between raised and lowered index positions, both for spatial and internal indices, since due to linearization these indices will be raised and lowered by Euclidean metrics $\delta_{ij}$ and $\delta_{ae}$.  For the same reason, these two types of indices will sometimes be interchangeable since they appear on equal footing at the linearized level.  Additionally, we will use the Einstein summation convention throughout this paper.  So for example, we have $a_{jj}=a^c_c=a^1_1+a^2_2+a^3_3$ and so on and so forth. 

\section{Initial value constraints about Petrov Type O}

For spacetimes of Petrov Type O we have $(\Psi_{ae})_o=-{3 \over \Lambda}\delta_{ae}$ with three equal eigenvalues and three linearly independent eigenvectors.  It is straightforward to see that this solves the constraints (\ref{VALUE}) for all connections $A^a_i$.  Indeed, replacing $\Psi_{ae}$ with $\delta_{ae}$ in the Gauss' law constraint yields\footnote{The Gauss' law constraint is simply the contraction of the spatial part of (\ref{FIRST2}) with a magnetic field $B^i_e$, yielding a covariant divergence on the internal index $e$.}
\begin{eqnarray}
\label{THEACTION71}
B^i_eD_i\delta_{ae}=B^i_e\partial_i\delta_{ae}+A^b_iB^i_e\bigl(f_{abf}\delta_{ge}+f_{ebg}\delta_{af}\bigr)\delta_{fg}\nonumber\\
=B^i_e\partial_i\delta_{ae}+A^b_iB^i_e(f_{abe}+f_{eba})=0~\forall{A}^a_i
\end{eqnarray}
\noindent
due to antisymmetry of the structure constants, and $\delta_{fg}$ being a numerical constant.  Replacing $\Psi_{ae}$ 
with $\delta_{ae}$ in the diffeomorphism constraint yields
\begin{eqnarray}
\label{YESDIR}
\epsilon_{ijk}B^j_aB^k_e\delta_{ae}=0~\forall{A}^a_i,
\end{eqnarray}
\noindent
due to the antisymmetry of the epsilon symbol.  Also, the Hamiltonian constraint
\begin{eqnarray}
\label{NODIR}
\hbox{tr}(\Psi^{-1})_o=-{\Lambda \over 3}\hbox{tr}(\delta_{ae})^{-1}=-\Lambda
\end{eqnarray}
\noindent
is identically satisfied for this choice.  In this paper we will use the instanton representation method to construct gravitational wave solutions by linearization about a particular Type O spacetime.  The graviton solution then will be defined by the Ansatz
\begin{equation}
\label{KODD}
\Psi_{ae}=-\bigl({3 \over \Lambda}\delta_{ae}+\epsilon_{ae}\bigr)
\end{equation}
\noindent
where $\epsilon_{ae}$ parametrizes the fluctuations about $(\Psi_{ae})_o$.  We will now expand the constraints (\ref{VALUE}) using equation (\ref{KODD}).  Note that the $(\Psi_{ae})_o$ part of (\ref{KODD}) becomes annihilated for the constraints linear 
in $\Psi_{ae}$.  So for more general $\Psi_{ae}$ for the Gauss' law constraint we have
\begin{eqnarray}
\label{KODD3}
B^i_eD_i\Psi_{ae}=B^i_eD_i\epsilon_{ae}=0,
\end{eqnarray}
\noindent
and for the diffeomorphism constraint we have
\begin{eqnarray}
\label{KODD5}
H_i=\epsilon_{ijk}B^j_aB^k_e\Psi_{ae}=\epsilon_{ijk}B^j_aB^k_e\epsilon_{ae}=0.
\end{eqnarray}
\noindent
For the Hamiltonian constraint we need the trace of the inverse of (\ref{KODD}), whose inverse is given by
\begin{eqnarray}
\label{INVERSEE}
(\Psi^{-1})^{ae}=-{\Lambda \over 3}\bigl(\delta_{ae}-{\Lambda \over 3}\epsilon_{ae}+\dots\bigr),
\end{eqnarray}
\noindent
where the dots represent all terms of second order in $\epsilon_{ae}$ and higher.  Taking the trace of (\ref{INVERSEE}), then we can write the constraints as the following system of seven equations in nine unknowns
\begin{eqnarray}
\label{CONSTRAINTTT}
\epsilon_{ijk}B^j_aB^k_e\epsilon_{ae}=0;~~
B^i_eD_i\{\epsilon_{ae}\}=0;~~{{\Lambda^2} \over 9}\hbox{tr}\epsilon+\dots=0.
\end{eqnarray}
\noindent
Note that the Gauss' law and diffeomorphism constraints are independent of $\Lambda$, since these constraints are 
linear in $\epsilon_{ae}$.  For the Hamiltonian constraint, an imprint of $\Lambda$ remains upon expansion due to the nonlinearity of the constraint.  This can be seen as the imprint of the Petrov Type O background, which interacts with the fluctuations.  

\section{Spatial 3-metric from the linearized constraints}

Having expanded $\Psi_{ae}$ in (\ref{KODD}) about a Type O solution, we will now linearize the constraints about this solution by taking $\epsilon_{ae}$ to be small.  First we will neglect all terms of second order and higher in $\epsilon_{ae}$, which reduces (\ref{CONSTRAINTTT}) to 
\begin{eqnarray}
\label{LINEARIZED}
\epsilon_{ijk}B^j_aB^k_e\epsilon_{ae}=0;~~
B^i_eD_i\{\epsilon_{ae}\}=0;~~
\hbox{tr}\epsilon=0.
\end{eqnarray}
\noindent
Next, we will linearize the connection $A^a_i$ about a reference background connection $\alpha^a_i$ 
\begin{eqnarray}
\label{NOWEXPAND}
A^a_i=\alpha^a_i+a^a_i,
\end{eqnarray}
\noindent
where $\vert{a}^a_i\vert<<\alpha^a_i$.  The linearized magnetic field for (\ref{NOWEXPAND}) is given by
\begin{eqnarray}
\label{NOWEXPAND1}
B^i_a=\epsilon^{ijk}\partial_j(\alpha^a_k+a^a_k)+{1 \over 2}\epsilon^{ijk}f^{abc}(\alpha^b_j+a^b_j)(\alpha^c_k+a^c_k)\nonumber\\
=\beta^i_a[\alpha]+\epsilon^{ijk}(\partial_ja^a_k+f^{abc}\alpha^b_ja^c_k)+O(a^2)
\end{eqnarray}
\noindent
where $\beta^i_a[\alpha]$ is the magnetic field of $\alpha^a_i$.  While any background connection $\alpha^a_i$ will suffice, a straightforward choice which as we will see clearly elucidates the physical content of the theory is a reference 
connection $\alpha^a_i=\delta^a_i\alpha$, where $\alpha=\alpha(t)$ is spatially homogeneous and depends only on time.  Then we have
\begin{eqnarray}
\label{NOWEXPAND2}
B^i_a=\alpha^2\Bigl(\delta^i_a+{1 \over \alpha}\bigl(\delta^i_a\hbox{tr}a-a^i_a\bigr)
+{1 \over {\alpha^2}}\epsilon^{ijk}\partial_ja^a_k\Bigr);~~A^a_iB^i_e=\delta_{ae}\alpha^3+\dots,
\end{eqnarray}
\noindent
where the dots signify all higher order terms.  Since the constraints (\ref{LINEARIZED}) are already of linear 
order in $\epsilon_{ae}$, then it suffices to retain only the zeroth order terms involving $A^a_i$ in order to complete the linearization.  Hence the linearized diffeomorphism constraint is given by
\begin{eqnarray}
\label{NOWEXPAND4}
H_i=\epsilon_{ijk}(\alpha^4\delta^j_a\delta^k_e)\epsilon_{ae}=\alpha^4\epsilon_{iae}\epsilon_{ae}=0,
\end{eqnarray}
\noindent
which implies that $\epsilon_{ae}=\epsilon_{ea}$ must be symmetric.  The Hamiltonian constraint to linearized order is given by
\begin{eqnarray}
\label{NOWEXPAND5}
\hbox{tr}\epsilon=0,
\end{eqnarray}
\noindent
which states that $\epsilon_{ae}$ is traceless to this order.  For the Gauss' law constraint we have
\begin{eqnarray}
\label{NOWEXPAND6}
G_a=\alpha^2\delta^i_e\partial_i\epsilon_{ae}+\alpha^3\delta_{be}\bigl(f_{abf}\delta_{ge}+f_{ebg}\delta_{af}\bigr)\epsilon_{fg}\nonumber\\
=\alpha^2\partial_e\epsilon_{ae}+\alpha^3f_{agf}\epsilon_{fg}=0.
\end{eqnarray}
\noindent
The second term on the right hand side of (\ref{NOWEXPAND6}) vanishes since $\epsilon_{ae}$ is symmetric from (\ref{NOWEXPAND4}), and the Gauss' law constraint reduces to
\begin{eqnarray}
\label{NOWEXPAND7}
\partial_e\epsilon_{ae}=0,
\end{eqnarray}
\noindent
which states that $\epsilon_{ae}$ is transverse.  So upon implementation of the linearized constraints $\epsilon_{ae}$ is symmetric, traceless and transverse, which means that it is a spin two field.\par
\indent

\subsection{Spatial 3-metric from the constraints}

The next step in the instanton representation method is now to compute the spatial 3-metric from the solution to the initial value constraints
\begin{eqnarray}
\label{INITIAL}
(h_{ij})_{Constraints}=(\hbox{det}\Psi)(\Psi^{-1}\Psi^{-1})^{ae}(B^{-1})^a_i(B^{-1})^e_j(\hbox{det}B)
\end{eqnarray}
\noindent
to linear order in $\epsilon_{ae}$ and $a^a_i$.  To keep organized let us first compute the ingredients of (\ref{INITIAL}).  The matrix $\Psi_{ae}$ is already of linear order as evident from (\ref{KODD}), repeated here for completeness
\begin{eqnarray}
\label{INITIALONE}
\Psi_{ae}=-{1 \over k}\bigl(\delta_{ae}+k\epsilon_{ae}\bigr);~~k={\Lambda \over 3},
\end{eqnarray}
\noindent
where $\epsilon_{ae}$ satisfies the conditions for a spin 2 field
\begin{eqnarray}
\label{MOTIVATED}
\epsilon_{dae}\epsilon_{ae}=0;~~\partial^e\epsilon_{ae}=0;~~\hbox{tr}\epsilon=0.
\end{eqnarray}
\noindent
The square of the inverse (\ref{INITIALONE}) and the determinant to linear order in $\epsilon_{ae}$ are given by
\begin{eqnarray}
\label{INITIALTWO}
(\Psi^{-1}\Psi^{-1})^{ae}=k^2\bigl(\delta_{ae}-2k\epsilon_{ae}\bigr);~~(\hbox{det}\Psi)=-{1 \over {k^3}}(1+k\hbox{tr}\epsilon)=-{1 \over {k^3}},
\end{eqnarray}
\noindent
where we have used the tracelessness of $\epsilon_{ae}$ from (\ref{MOTIVATED}).  The linearized determinant of the magnetic 
field from (\ref{NOWEXPAND2}) is given by
\begin{eqnarray}
\label{LINEARIZEDDETERM}
\hbox{det}B=\alpha^6\Bigl(1+{2 \over \alpha}\hbox{tr}a+{1 \over {\alpha^2}}\partial_j(\epsilon^{ijk}a^i_k)\Bigr),
\end{eqnarray}
\noindent
and the linearized inverse is given by
\begin{eqnarray}
\label{HOOFT16}
(B^{-1})^a_i={1 \over {\alpha^2}}\Bigl(\delta^a_i+{1 \over \alpha}\bigl(a^a_i-\delta^a_i(\hbox{tr}a)\bigr)-{1 \over {\alpha^2}}\epsilon^{imk}\partial_ma^a_k\Bigr).
\end{eqnarray}
\noindent
Given that $\epsilon_{ae}$ is symmetric, transverse and traceless on account of (\ref{MOTIVATED}), it seems natural that the connection perturbation $a^a_i$ should also exhibit these properties.  Let us impose the conditions\footnote{The linearized initial value constraints (\ref{MOTIVATED}) constrain $\epsilon_{ae}$ and not $a^a_i$, therefore (\ref{ASSUMPTIONS}) can be regarded as a gauge-fixing choice of the connection $A^a_i$.  We will see later in this paper that (\ref{ASSUMPTIONS}) is self-consistent and consistent with (\ref{MOTIVATED}) and with the equations of motion, which provides justification for this choice.} 
\begin{eqnarray}
\label{ASSUMPTIONS}
\epsilon_{ijk}a^j_k=0;~~\hbox{tr}a=a^c_c=0;~~\partial^ka^j_k=0.
\end{eqnarray}
\noindent
Then the spatial and the internal indices of $a^a_i$ are now on the same footing.  Equations (\ref{HOOFT16}) and 
(\ref{LINEARIZEDDETERM}) simplify to
\begin{eqnarray}
\label{SIMPLIFYTO}
(B^{-1})^a_i=\alpha^{-2}\bigl(\delta^a_i+\alpha^{-1}a^a_i-\alpha^{-2}\epsilon^{imk}\partial_ma^a_k\bigr);~~\hbox{det}B=\alpha^6.
\end{eqnarray}
Substituting (\ref{INITIALTWO}) and (\ref{SIMPLIFYTO}) into (\ref{INITIAL}), we get the spatial 3-metric to linearized order based on the constraint solutions
\begin{eqnarray}
\label{INITIALTHREE}
(h_{ij})_{Constraints}=-{{\alpha^2} \over k}\Bigl(\delta_{ij}-2k\epsilon_{(ij)}+{2 \over \alpha}a_{(ij)}-{2 \over {\alpha^2}}\epsilon^{(imn}\partial_ma^{j)}_n\Bigr).
\end{eqnarray}
\noindent
We have shown that the initial value constraints at the linearized level confer the massless spin 2 
polarizations on $\epsilon_{ae}$, and we have imposed associated gauge-fixing condition (\ref{ASSUMPTIONS}) by hand on the connection perturbation $a^a_i$.

\section{Spatial 3-metric from Hodge duality condition}

Equation (\ref{INITIALTHREE}) depends both on $\epsilon_{ae}$ and $a^a_i$ restricted to a particular spatial 
hypersurface $\Sigma$, and the linearized constraints (\ref{MOTIVATED}) are insufficient to prescribe their time evolution.  To make progress we must next determine 3-metric based on the Hodge duality condition, given by
\begin{eqnarray}
\label{HODEGE}
(h_{ij})_{Hodge}=-{{N^2} \over c}c_{(ij)};~~c_{ij}=F^a_{0i}(B^{-1})^a_j.
\end{eqnarray}
\noindent
To keep organized we will first compute the ingredients of (\ref{HODEGE}).  The temporal component of the curvature is given by
\begin{eqnarray}
\label{MAGNET}
F^a_{0i}=\dot{A}^a_i-D_iA^a_0=\dot{A}^a_i-\bigl(\partial_iA^a_0+f^{abc}A^b_iA^c_0\bigr).
\end{eqnarray}
\noindent
In the initial value constraints we have used only a spatial 3-dimensional connection $A^a_i$.  For the Hodge duality condition we will use a linearized Ansatz for the 4-dimensional connection $A^a_{\mu}$ given by
\begin{eqnarray}
\label{MAGNET1}
A^a_{\mu}=\delta^a_{\mu}\alpha+a^a_{\mu},
\end{eqnarray}
\noindent
where we have defined $\delta^a_0=0$ and $\vert{a}^a_{\mu}\vert<<\alpha$.  Let us now compute $(h_{ij})_{Hodge}$ to linearized order, using $a^i_i=0$ from (\ref{ASSUMPTIONS}).  Equation (\ref{MAGNET}) to first order is 
\begin{eqnarray}
\label{HOOFT161}
F^a_{0i}=\dot{\alpha}\Bigl(\delta^a_i+{{\dot{a}^a_i} \over {\dot{\alpha}}}-{1 \over {\dot{\alpha}}}\partial_in^a-{\alpha \over {\dot{\alpha}}}f^{aic}n^c\Bigr),
\end{eqnarray}
\noindent
where we have defined $n^a\equiv{A}^a_0$ as the temporal component of $A^a_{\mu}$ and we have treated $n^a$ as small similarly to $a^a_i$.  Making use of (\ref{ASSUMPTIONS}) and the inverse magnetic field (\ref{SIMPLIFYTO}), then the following relation ensues to linearized order
\begin{eqnarray}
\label{HOOFT162}
c_{ij}=F^a_{0i}(B^{-1})^a_j={{\dot{\alpha}} \over {\alpha^2}}\Bigl(\delta_{ij}+{1 \over \alpha}a_{ij}\nonumber\\
-{1 \over {\alpha^2}}\epsilon^{jmn}\partial_ma_{in}
+{{\dot{a}_{ij}} \over {\dot{\alpha}}}-{1 \over {\dot{\alpha}}}\partial_in_j+{\alpha \over {\dot{\alpha}}}f_{ijk}n_k\Bigr).
\end{eqnarray}
\noindent
The symmetric and antisymmetric parts of (\ref{HOOFT162}) are given by
\begin{eqnarray}
\label{SYMANTPART}
c_{(ij)}={{\dot{\alpha}} \over {\alpha^2}}\Bigl(\delta_{ij}+{1 \over \alpha}a_{ij}+{1 \over {\dot{\alpha}}}\dot{a}_{ij}-{1 \over {\alpha^2}}\epsilon^{(jmk}\partial_ma_{ki)}-{1 \over {\dot{\alpha}}}\partial_{(i}n_{j)}\Bigr);\nonumber\\
c_{[ij]}={{\dot{\alpha}} \over {\alpha^2}}\Bigl(-{1 \over {\alpha^2}}\epsilon^{[jmk}\partial_ma_{ki]}-{1 \over {\dot{\alpha}}}\partial_{[i}n_{j]}-{\alpha \over {\dot{\alpha}}}f_{ijk}n^k\Bigr),
\end{eqnarray}
\noindent
where it is understood that $a_{ij}$ is already symmetric on account of (\ref{ASSUMPTIONS}).  The determinant of the symmetric part if $c_{ij}$ is given by
\begin{eqnarray}
\label{WITHDETERM}
c=\hbox{det}(c_{(ij)})=\Bigl({{\dot{\alpha}} \over {\alpha^2}}\Bigr)^3\bigl(1-{1 \over {\dot{\alpha}}}\partial_mn^m\bigr)
\end{eqnarray}
\noindent
where we have used the tracelessness and symmetry of $a^a_i$ from (\ref{ASSUMPTIONS}).  Substituting (\ref{SYMANTPART}) into (\ref{HODEGE}), we get the 3-metric from the Hodge duality condition 
\begin{eqnarray}
\label{INITIALFOUR}
(h_{ij})_{Hodge}=-N^2\Bigl({{\alpha^2} \over {\dot{\alpha}}}\Bigr)^2\bigl(1-{1 \over {\dot{\alpha}}}\partial_mn^m\bigr)^{-1}\Bigl(\delta_{ij}+{1 \over \alpha}a_{ij}\nonumber\\
-{1 \over {\alpha^2}}\epsilon^{(imn}\partial_ma_{j)n}
+{{\dot{a}_{ij}} \over {\dot{\alpha}}}-{1 \over {\dot{\alpha}}}\partial_{(i}n_{j)}\Bigr).
\end{eqnarray}
\noindent

\subsection{Consistency condition on the background solution}

We have computed the 3-metric $h_{ij}$ based upon two separate criteria.  As a consistency condition we must require that
\begin{eqnarray}
\label{ASACONSISTENT}
(h_{ij})_{Hodge}=(h_{ij})_{Constraints},
\end{eqnarray}
\noindent
which leads to the equation
\begin{eqnarray}
\label{ASACONSISTENT1}
-N^2\Bigl({{\alpha^2} \over {\dot{\alpha}}}\Bigr)^2\bigl(1-{1 \over {\dot{\alpha}}}\partial_mn^m\bigr)^{-1}\Bigl(\delta_{ij}+{1 \over \alpha}a_{ij}-{1 \over {\alpha^2}}\epsilon^{(imn}\partial_ma_{nj)}\nonumber\\
+{1 \over {\dot{\alpha}}}\dot{a}_{ij}-{1 \over {\dot{\alpha}}}\partial_{(i}n_{j)}\Bigr)
=-\Bigl({{\alpha^2} \over k}\Bigr)\Bigl(\delta_{ij}-2k\epsilon_{ij}+{2 \over \alpha}a_{ij}-{2 \over {\alpha^2}}\epsilon^{(imn}\partial_ma_{nj)}\Bigr).
\end{eqnarray}
\noindent
Equation (\ref{ASACONSISTENT1}) will put strong constraints on the form of the metric solution.  To start with, we can set the pre-factors in (\ref{ASACONSISTENT1}) equal to each other 
\begin{eqnarray}
\label{INITIALFIVE}
{{\alpha^2} \over k}=\bigl(1-{1 \over {\dot{\alpha}}}\partial_mn^m\bigr)^{-1}N^2{{\alpha^4} \over {\dot{\alpha}^2}},
\end{eqnarray}
\noindent
which will fix the background solution.  Recall that $\alpha=\alpha(t)$ by supposition is spatially homogeneous and depends only on time.  In order for (\ref{INITIALFIVE}) to be consistent, then the lapse function $N$ must be chosen such that its spatial dependence cancels out any spatial dependence due to $\partial_in^i$.  We will choose $\partial_in^i=0$ for simplicity,\footnote{In due course we will show that this choice is not arbitrary, but is actually a consistency condition which follows from the equations of motion.} and choose $N=N(t)$ to depend only on time.  Given these conditions, then equation (\ref{INITIALFIVE}) is a first order linear differential equation for $\alpha(t)$, which integrates directly to
\begin{eqnarray}
\label{INITIALSIX}
\alpha(t)=\alpha(0)\hbox{exp}\Bigl[\sqrt{{\Lambda \over 3}}\int^t_0N(t^{\prime})dt^{\prime}\Bigr].
\end{eqnarray}
\noindent
The background 3-metric is given by (\ref{INITIAL}) with 
$(\Psi_{ae})_o=-{3 \over \Lambda}\delta_{ae}$ and $(A^a_i)_o=\delta^a_i\alpha$, which is
\begin{eqnarray}
\label{BACKMETRIC}
(h_{ij})_o=-\delta_{ij}{{3\alpha_0^2} \over \Lambda}\hbox{exp}\Bigl[2\sqrt{{\Lambda \over 3}}\int^t_0N(t^{\prime})dt^{\prime}\Bigr].
\end{eqnarray}
\noindent
Reality conditions on the background solution dictate that for $\Lambda>0$, $N=1$ is a suitable choice of lapse function.  Therefore for $\Lambda<0$, then $N=i$ is suitable.  For $\alpha_0$ real we have a Euclidean signature metric, whereas for $\alpha_0$ pure imaginary we have Lorentizian signature.\footnote{Observe that it is the initial 
data $\alpha_0$ and not the lapse function $N$ which determines the signature of background spacetime.  Nevertheless, we will restrict ourselves to real $\alpha_0$ in order that the metric perturbations be real.  This will limit our result to gravitons propagating on Euclidean DeSitter spacetime.}  Choosing $\alpha_0=\sqrt{\Lambda \over 3}$, the solution for the background metric is this gauge is given by 
\begin{eqnarray}
\label{BACKGROUNDMETRIC}
(ds^2)_o=-\bigl(dt^2\pm{e}^{2\sqrt{{\Lambda \over 3}}t}(dx^2+dy^2+dz^2)\bigr),
\end{eqnarray}
\noindent
which is the metric for an inflating de Sitter background.  This provides the physical interpretation that the length scale associated with $\alpha_0$, the initial value of the background connection, is the DeSitter radius $l_0=\sqrt{{3 \over \Lambda}}$. 
\noindent

\subsection{Consistency condition on the perturbation}

Having determined the background solution from consistency of (\ref{ASACONSISTENT}), we will now follow suit for the linearized perturbation of $h_{ij}$.  Equality of the terms of (\ref{ASACONSISTENT1}) in large round brackets leads to the following first order differential equation for the connection perturbation 
\begin{eqnarray}
\label{INITIALSEVEN}
\dot{a}_{ij}=\partial_{(i}n_{j)}-2k\dot{\alpha}\epsilon_{ij}+{{\dot{\alpha}} \over \alpha}a_{ij}
-{{\dot{\alpha}} \over {\alpha^2}}\epsilon^{(imn}\partial_ma_{nj)}.
\end{eqnarray}
\noindent
Prior to proceeding we must check that (\ref{INITIALSEVEN}) is consistent with (\ref{MOTIVATED}) and (\ref{ASSUMPTIONS}).  First, note that the antisymmetric part of (\ref{INITIALSEVEN}) is zero since the equation is already symmetric in $ij$.  The trace of (\ref{INITIALSEVEN}) implies that
\begin{eqnarray}
\label{THETRACE}
\partial_in^i=0,
\end{eqnarray}
\noindent
where we have used that $\epsilon_{ij},a_{ij}$ are symmetric and traceless.\footnote{Equation (\ref{THETRACE}) is the aforementioned consistency condition 
on (\ref{INITIALFIVE}) which requires that the background 3-metric be spatially homogeneous for spatially homogeneous lapse $N$.}  Lastly, we must show that (\ref{INITIALSEVEN}) is transverse.  Acting on (\ref{INITIALSEVEN}) with $\partial^i$ and using $\partial^i\epsilon_{ij}=\partial^ia_{ij}=0$, we have
\begin{eqnarray}
\label{THETRACE1}
\partial^i(\partial_in_j+\partial_jn_i)={{\dot{\alpha}} \over {\alpha^2}}\bigl[\epsilon^{imn}\partial_i\partial_ma_{nj}+\epsilon^{jmn}\partial_i\partial_ma_{ni}\bigr].
\end{eqnarray}
\noindent
The terms in square brackets in (\ref{THETRACE1}) vanish due to antisymmetry of $\epsilon^{imn}$ and the transversality of $a_{nj}$.  Using $\partial^in_i=0$ from (\ref{THETRACE}), then this implies
\begin{eqnarray}
\label{THETRACE2}
\partial^2n_j=0.
\end{eqnarray}
\noindent
The resulting consistency condition on (\ref{INITIALSEVEN}) is that the temporal connection component of $A^a_0=n^a$ must satisfy the Laplace equation.\par  
\indent
We can now compute the shift vector from the antisymmetric part of $c_{ij}$ from (\ref{SYMANTPART}).  The shift vector $N^i$ is given by
\begin{eqnarray}
\label{THESHIFTVECTOR}
N^k=-{1 \over 2}\epsilon^{kij}c_{ij}\nonumber\\
=-{{\dot{\alpha}} \over {\alpha^2}}\epsilon^{kij}\Bigl({1 \over {\alpha^2}}\epsilon^{imn}\partial_ma_{nj}-{1 \over {\alpha^2}}\epsilon^{jmn}\partial_ma_{ni}\Bigr)
-{1 \over {\alpha^2}}\epsilon^{kij}\partial_in_j+{{\dot{\alpha}} \over {\alpha^2}}n^k.
\end{eqnarray}
\noindent
Applying epsilon tensor identities to the terms in round brackets in (\ref{THESHIFTVECTOR}) 
\begin{eqnarray}
\label{THESHIFTVECTOR1}
\bigl(\delta^{mj}\delta^{nk}-\delta^{mk}\delta^{nj}\bigr)\partial_ma_{nj}-\bigl(\delta^{mk}\delta^{ni}-\delta^{mi}\delta^{nk}\bigr)\partial_ma_{ni}
=2\partial_ja_{kj}-2\partial_ka_{jj}=0,
\end{eqnarray}
\noindent
we see that these terms vanish on account of the transversality and tracelessness of $a_{ij}$.  Therefore (\ref{THESHIFTVECTOR}) reduces to
\begin{eqnarray}
\label{THESHIFTVECTOR2}
N^k=-{1 \over {\alpha^2}}\epsilon^{kij}\partial_i{n}_j+{{\dot{\alpha}} \over {\alpha^2}}{n}^k.
\end{eqnarray}
\noindent
There is a one-to-one correlation between temporal connection components $n^i=A^i_0$ and the shift vector $N^i$, which are gauge degrees of freedom respectively in the Yang--Mills and the metric formulations of gravity.\par
\indent
Having verified the consistency of (\ref{MOTIVATED}) and (\ref{ASSUMPTIONS}) with the Hodge duality condition, we can write (\ref{INITIALSEVEN}) as the differential equation
\begin{eqnarray}
\label{THESHIFTVECTOR3}
\dot{a}_{ij}=\partial_{(i}n_{j)}+\bigl(c_1\delta^k_i\delta^l_j+c_2\eta^{kl}_{ij}\bigr)a_{kl}+c_3\epsilon_{ij},
\end{eqnarray}
\noindent
where we have defined
\begin{eqnarray}
\label{THESHIFTVECTOR4}
c_1(t)={{\dot{\alpha}} \over \alpha};~~c_2(t)=-{{\dot{\alpha}} \over {\alpha^2}};~~c_3(t)=-2k\dot{\alpha};~~
\eta_{(ij)}^{kl}=\epsilon^{(iml}\delta^k_{j)}\partial_m.
\end{eqnarray}
\noindent
Note that $\eta_{ij}^{kl}$ is a differential operator.  Equation (\ref{THESHIFTVECTOR3}) is a linear first order differential equation for the connection perturbation $a_{ij}$ totally consistent with (\ref{ASSUMPTIONS}), but also 
involves $\epsilon_{ij}$.  To integrate this equation, we need to know the explicit time dependence of $\epsilon_{ij}$.

\section{Time evolution of the deviation matrix $\epsilon_{ae}$}

Equation (\ref{THESHIFTVECTOR3}) is a linear first order evolution equation for the connection perturbation $a_{ij}$, which arose from (\ref{ASACONSISTENT}).  To integrate (\ref{THESHIFTVECTOR3}) we need to know the time 
dependence of $\epsilon_{ae}$, which cannot be determined from (\ref{RECALLL}) and (\ref{MOTIVATED}).  This is where equation (\ref{RECALL2}) comes into play.  The temporal part 
of (\ref{RECALL2}) has already been used via Gauss' law to conclude that $\epsilon_{ae}$ is transverse, leaving remaining the spatial parts.  Using $\epsilon^{0ijk}=\epsilon^{ijk}$, as well as the 
definition $B^i_e={1 \over 2}\epsilon^{ijk}F^e_{jk}$, then the $\mu=i$ components of (\ref{RECALL2}) yield the equations
\begin{eqnarray}
\label{RECALL3}
-2B^i_eD_0\Psi_{ae}+2\epsilon^{ijk}F^e_{ok}D_j\Psi_{ae}=0.
\end{eqnarray}
\noindent
As a consistency condition we will first verify that the linearization of (\ref{RECALL3}) is consistent with (\ref{MOTIVATED}).  To perform the linearization, it will be convenient to transfer the magnetic field to the right hand side of (\ref{RECALL3}), yielding
\begin{eqnarray}
\label{RECALL4}
D_0\Psi_{ae}=\epsilon^{ijk}(B^{-1})^e_iF^f_{0k}D_j\Psi_{af}.
\end{eqnarray}
\noindent
We will now substitute (\ref{KODD}) and (\ref{MAGNET1}) into (\ref{RECALL4}).  First note that 
the $(\Psi_{ae})_o\propto\delta_{ae}$ part is annihilated by the gauge covariant derivatives (\ref{FIRST2}), since $\alpha^b_0=0$ and due to antisymmetry of the structure constants.  Then since (\ref{RECALL4}) is already linear 
in $\epsilon_{ae}$, then it suffices to expand (\ref{RECALL4}) to zeroth order in $a^a_i$ in order to carry out the linearization, which yields
\begin{eqnarray}
\label{RECALL6}
\dot{\epsilon}_{ae}=\epsilon^{ijk}(\delta^e_i\alpha^{-2})(\delta^f_k\dot{\alpha})\partial_j\epsilon_{af}.
\end{eqnarray}
\noindent
Since $\epsilon_{ae}$ is symmetric, then this yields implies the equation
\begin{eqnarray}
\label{RECALL7}
\dot{\epsilon}_{ae}=c_2\epsilon^{(ejf}\partial_j\epsilon_{a)f}
\end{eqnarray}
\noindent
with $c_2$ as in (\ref{THESHIFTVECTOR4}).  To linearized order, there is no information from $a_{ij}$ contained in (\ref{RECALL7}), which is the same situation as for the linearized initial value constraints (\ref{MOTIVATED}).  Before proceeding with the solution, let us check the consistency of (\ref{RECALL7}) with (\ref{MOTIVATED}).  The trace of (\ref{RECALL7}) implies
\begin{eqnarray}
\label{RECALL8}
\dot{\epsilon}_{aa}=c_2\epsilon^{ajf}\partial_j\epsilon_{af}=0
\end{eqnarray}
\noindent
since $\epsilon_{af}$ is symmetric.  Acting on (\ref{RECALL7}) with $\partial^e$ yields
\begin{eqnarray}
\label{RECALL9}
\partial^e\dot{\epsilon}_{ae}=c_2\epsilon^{ejf}\partial_e\partial_j\epsilon_{af}=0,
\end{eqnarray}
\noindent
which demonstrates transversality on the second index.  Acting on (\ref{RECALL7}) with $\partial^a$ yields
\begin{eqnarray}
\label{RECALL10}
\partial^a\dot{\epsilon}_{ae}=c_2\epsilon^{ejf}\partial_j\partial^a\epsilon_{af}=0
\end{eqnarray}
\noindent
which is consistent with transversality on the first index.  Lastly, we must prove consistency with the 
symmetry of $\epsilon_{ae}$.  The antisymmetric part of (\ref{RECALL7}) is given by
\begin{eqnarray}
\label{RECALL11}
\epsilon_{dae}\dot{\epsilon}_{ae}=c_2\epsilon_{eda}\epsilon^{ejf}\partial_j\epsilon_{af}
=c_2\bigl(\delta^j_d\delta^f_a-\delta^j_a\delta^f_d\bigr)\partial_j\epsilon_{af}
=c_2(\partial_d\epsilon_{aa}-\partial_a\epsilon_{ad})=0,
\end{eqnarray}
\noindent
where we have used the Einstein summation convention and the tracelessness and transversality of $\epsilon_{ae}$.  The result is that (\ref{RECALL2}) is consistent with the initial value constraints.\par
\indent
Equation (\ref{RECALL7}) can be written as
\begin{eqnarray}
\label{RECALL12}
\dot{\epsilon}_{ae}=c_2\eta_{ae}^{bf}\epsilon_{bf},
\end{eqnarray}
\noindent
with solution (see Appendix A for the derivation)
\begin{eqnarray}
\label{RECALL13}
\epsilon_{ae}(x,t)=U_{ae}^{bf}[t,0]_{\eta}\epsilon_{bf}(x,0),
\end{eqnarray}
\noindent
where we have defined the time-ordered exponential $U$ by
\begin{eqnarray}
\label{RECALL14}
U_{ae}^{bf}[t,0]_{\eta}
=\hat{T}\Bigl\{\hbox{exp}\Bigl[\boldsymbol{\eta}\int^t_0dt^{\prime}c_2(t^{\prime})\Bigr]\Bigr\}^{bf}_{ae}.
\end{eqnarray}
\noindent
where $\hat{T}$ is the time-ordering operator and $\boldsymbol{\eta}$ is the operator matrix $\eta_{ae}^{bf}$.  So we have determined the time evolution of $\epsilon_{ae}$ from its initial value as a spin two field in accordance with the initial value constraints.  Note that the solution to the constraints is preserved for all time.  For restrictions on the initial data we must 
have $\epsilon_{bf}(x,0)\in{C}^{\infty}(\Sigma)$ so that all spatial derivatives coming down from the exponential exist and are well-defined.

\section{The connection and the spatial 3-metric}

Having determined the time evolution for $\epsilon_{ae}$ from an initial spatial hypersurface $\Sigma_0$, we can now find the metric as follows.  First substitute (\ref{RECALL3}) into (\ref{INITIALSEVEN}), which then leads to an explicit differential equation for $a_{ij}$.  Using similar time-ordered exponential techniques, we can find
\begin{eqnarray}
\label{RECALL15}
a_{ij}(x,t)=\int^t_0dt^{\prime}U_{ij}^{kl}[t^{\prime},0]\partial_{(k}n_{l)}(x,t^{\prime})
+U_{ij}^{kl}[t,0]_Pa_{kl}(x,0)\nonumber\\
+U_{ij}^{kl}[t,0]_P\int^{t^{\prime}}_0dt^{\prime\prime}U_{kl}^{mn}[t^{\prime},0]_{\eta}\epsilon_{mn}(x,0),
\end{eqnarray}
\noindent
where we have defined the time-ordered exponential operator (in analogy with (\ref{RECALL14}))
\begin{eqnarray}
\label{RECALL16}
U_{ae}[t,0]_P
=\hat{T}\Bigl\{\hbox{exp}\Bigl[\boldsymbol{P}\int^t_0dt^{\prime}c_2(t^{\prime})\Bigr]\Bigr\}^{bf}_{ae}.
\end{eqnarray}
\noindent
with the differential operator $P_{ij}^{kl}$ defined as
\begin{eqnarray}
\label{RECALL17}
P_{ij}^{kl}=c_1(t)\delta_i^k\delta_j^l+c_2(t)\eta_{ij}^{kl}.
\end{eqnarray}
\noindent
The main point is that (\ref{RECALL15}) consists of evolution operators acting on some linear combination of the initial data of the basic variables capturing the physical degrees of freedom, as well as the unphysical ones encoded in $n_i$.  Substituting (\ref{RECALL15}) and (\ref{RECALL13}) into (\ref{INITIALTHREE}), we can now construct the 3-metric in the general form
\begin{eqnarray}
\label{RECALL18}
h_{ij}(x,t)=-{{\alpha^2} \over k}\Bigl(\delta_{ij}+A_{ij}^{mn}(t,0)\partial_{(m}n_{n)}(x,0)\nonumber\\
+B_{ij}^{mn}(t,0)a_{mn}(x,0)+C_{ij}^{mn}(t,0)\epsilon_{mn}(x,0)\Bigr),
\end{eqnarray}
\noindent
where $A$, $B$ and $C$ consist of time-ordered evolution operators containing derivative operators, whose specific form we will not display here.  The point is that $h_{ij}$ has been reduced explicitly to a time evolution of the initial data $\epsilon_{mn}(x,0)$ and $a_{mn}(x,0)$ satisfying the initial value constraints and gauge conditions.  The quantity $\partial_{(m}n_{n)}$ can be seen as the Lie derivative of $h_{ij}$ along the vector $n^a=A^a_0$, which takes the interpretation of a spatial coordinate transformation encoding the gauge degrees of freedom.  The spin 2 nature of the metric is preserved for all 
time and is cleanly separated from the gauge degrees of freedom, namely the temporal connection components $A^a_0=n^a$ in the form of the shift vector $N^i$.  Combining (\ref{RECALL18}) with the shift vector (\ref{THESHIFTVECTOR}), this provides the spacetime metric for the gravitons about DeSitter spacetime.  One should expect to be able to apply a similar algorithm for expansion in gravitons about other exact General Relativity solutions.

\section{Summary}

The main result of this paper has been the application of the instanton representation method to the construction of graviton solutions for General Relativity, linearized about a Euclidean DeSitter spacetime background.  We have used this solution, which is well-known in the literature, as a testing ground for the method.  Starting from the initial value constraints combined with gauge-fixing conditions, we have obtained the spin 2 polarization for the basic fields.  This prescribed the physical degrees of freedom for gravity on an initial spatial hypersurface $\Sigma_0$.  Using the instanton representation equations of motion, we determined the evolution of these physical degrees of freedom in time.  The significance of this is that the spin 2 polarizations remain preserved as a consequence of the constraints' being consistent with the evolution equations, a feature which is explicit.  From these solutions we constructed the spatial 3-metric explicitly, depicting a neat separation of the physical from unphysical degrees of freedom.  The unphysical degrees of freedom are due to the temporal component of the $SO(3,C)$ connection $A^a_0=n^a$, which correlate directly to the shift vector $N^i$.  This provides a physical basis and interpretation for the role of the gauge connection and its relation to metric General Relativity.  Our next work will be to reconstruct minisuperspace solutions as a further testing arena, and then subsequently to generate new solutions to the Einstein equations which are not known in the literature.

\section{Appendix A}

We would like to solve the differential equation
\begin{eqnarray}
\label{WEWOULD}
{d \over {dt}}\epsilon_{ae}(x,t)=K(t)\eta^{bf}_{ae}\epsilon_{bf}(x,t).
\end{eqnarray}
\noindent
Integration of (\ref{WEWOULD}) from $0$ to $t$ yields
\begin{eqnarray}
\label{WEWOULD1}
\epsilon_{ae}(x,t)=\epsilon_{ae}(x,0)+\eta^{bf}_{ae}\int^t_0K(t^{\prime})\epsilon_{bf}(x,t^{\prime})dt^{\prime}.
\end{eqnarray}
\noindent
We have brought $\eta^{bf}_{ae}$ outside the integral, since it contains spatial derivative operators which commute with the time integration.  Equation (\ref{WEWOULD1}) can be iterated to
\begin{eqnarray}
\label{WEWOULD2}
\epsilon_{ae}(x,t)=\epsilon_{ae}(x,0)
+\int^t_0K(t_1)\eta^{a_1e_1}_{ae}\Bigl[\epsilon_{a_1e_1}(x,0)+\int^{t_1}_0K(t_2)\eta^{a_2e_2}_{ae}\epsilon_{a_2e_2}(x,t_2)dt_2\Bigr]dt_1.
\end{eqnarray}
\noindent
Continuing the iteration, this yields
\begin{eqnarray}
\label{WEWOULD3}
\biggl[\delta_a^b\delta_e^f+\eta^{bf}_{ae}\int^t_0dt_1K(t_1)+\eta_{ae}^{a_1e_1}\eta_{a_1e_1}^{bf}\int^t_0dt_1K(t_1)
\int^{t_1}_0dt_2K(t_2)\nonumber\\
+\eta_{ae}^{a_1e_1}\eta_{a_1e_1}^{a_2e_2}\eta^{bf}_{a_2e_2}\int^t_0dt_1K(t_1)\int^{t_2}_0dt_2K(t_2)\int^{t_2}_0dt_3K(t_3)
+\dots\biggr]\epsilon_{bf}(x,0).
\end{eqnarray}
\noindent
Analogy with the time-ordered Wick expansion in field theory signifies that we can make the upper limits of all time integrations the same and introduce a compensating factor, which yields
\begin{eqnarray}
\label{WEWOULD4}
\epsilon_{ae}(x,t)=\biggl[\delta_a^b\delta_e^f+\delta^b_{a_n}\delta^f_{e_n}\sum_{n=1}^{\infty}
\eta_{ae}^{a_1e_1}\eta_{a_1e_1}^{a_2e_2}\dots\eta_{a_{n-1}e_{n-1}}^{a_ne_n}
\hat{T}\Bigl\{\Bigl(\int_0^tdt^{\prime}K(t^{\prime})\Bigr)^n\Bigr\}\Bigr]\epsilon_{bf}(x,0)
\end{eqnarray}


\begin{thebibliography}{99}

\bibitem{EYOITA} {Eyo Eyo Ita III `Instanton representation of Plebanski gravity. Gravitational instantons from the classical formalism.'
Abraham Zelmanov Journal, 2011, volume 4, pages 36-71}

\bibitem{ASH2} {Abhay Ashtekar `New Hamiltonian formulation of general relativity'
Phys. Rev. D36(1987)1587}

\bibitem{EYOITA1} {Eyo Eyo Ita III `Instanton representation of Plebanski gravity. Application to Schwarzschild metrics.'
Abraham Zelmanov Journal, 2011, volume 4, pages 72-95}

\bibitem{CAP} {Richard Capovilla, Ted Jacobson, John Dell `General Relativity without the Metric'
Class. Quant. Grav. Vol 63, Number 21 (1989) 2325-2328}

\bibitem{MACCALLUM} {Hans Stephani, Dietrich Kramer, Maclcolm MacCallum, Cornelius Hoenselaers, and Eduard Herlt `Exact Solutions of Einstein's Field Equations'
Cambridge University Press}

\bibitem{PENROSERIND} {R. Penrose and W. Rindler `Spinors and space-time'
Cambridge Monographs in Mathematical Physics}

\bibitem{SPINCON} {Richard Capovilla, John Dell and Ted Jacobson `A pure spin-connection formulation of gravity'
Class. Quantum. Grav. 8(1991)59-73}

\end{thebibliography}
\end{document}